\documentclass[useAMS,usenatbib]{mn2e}

\usepackage{graphicx}
\usepackage{amssymb}
\usepackage{amsmath}
\usepackage{lscape}
\usepackage{url}

\def \HI {H\,{\sc{i}}}
\def \MHI {$M_{\mathrm{HI}}$}
\def \Msun {$\mathrm{M_{\odot}}$}
\def \Mbary {$M_{\mathrm{baryons}}$}
\def \kms {km\,s$^{-1}$}

\def \Mstellar {$M_{\ast}$}
\def \Lstellar {$L_{\ast}$}

\def \Htwo {H$_{2}$}
\def \fHI {$f_{\mathrm{HI}}$}

\title[Variation of galactic cold gas reservoirs]{Variation of
  galactic cold gas reservoirs with stellar mass}

\author[]
{Natasha Maddox\thanks{maddox@astron.nl}$^{1,2}$,  Kelley M. Hess$^{1,3,2}$,
Danail Obreschkow$^{4}$, M.~J. Jarvis$^{5,6}$, 
\newauthor S.-L. Blyth$^1$
\vspace*{6pt}\\
$^1$Astrophysics, Cosmology and Gravity Centre (ACGC), Astronomy
Department, University of Cape Town, \\
Private Bag X3, 7701 Rondebosch, Republic of South Africa\\
$^2$Netherlands Institute for Radio Astronomy (ASTRON), PO Box 2,
7990 AA Dwingeloo, The Netherlands\\
$^3$Kapteyn Astronomical Institute, University of Groningen, PO Box
800, 9700 AV Groningen, The Netherlands\\
$^4$International Centre for Radio Astronomy Research (ICRAR), M468,
University of Western Australia, \\
35 Stirling Hwy, Crawley, WA 6009, Australia\\
$^5$Oxford Astrophysics, Denys Wilkinson Building,
University of Oxford, Keble Rd, Oxford, OX1 3RH, UK\\
$^6$Physics Department, University of the Western Cape,
Cape Town, 7535, Republic of South Africa\\
}

\begin{document}


\maketitle

\begin{abstract}

The stellar and neutral hydrogen (\HI) mass functions at $z\sim 0$ are
fundamental benchmarks for current models of galaxy evolution. A
natural extension of these benchmarks is the two-dimensional
distribution of galaxies in the plane spanned by stellar and \HI\
mass, which provides a more stringent test of simulations, as it
requires the \HI\ to be located in galaxies of the correct stellar
mass. Combining \HI\ data from the ALFALFA survey, with optical data
from SDSS, we find a distinct envelope in the \HI-to-stellar mass
distribution, corresponding to an upper limit in the \HI\ fraction
that varies monotonically over five orders of magnitude in stellar
mass. This upper envelope in \HI\ fraction does not favour the
existence of a significant population of dark galaxies with large amounts of
gas but no corresponding stellar population. The envelope shows a
break at a stellar mass of $\sim 10^9$\,\Msun, which is not reproduced
by modern models of galaxy populations tracing both stellar and gas
masses. The discrepancy between observations and models suggests a
mass dependence in gas storage and consumption missing in current galaxy evolution
prescriptions. The break coincides with the transition from galaxies
with predominantly irregular morphology at low masses to regular disks
at high masses, as well as the transition from cold to hot accretion
of gas in simulations.

\end{abstract}


\section{Introduction}\label{sec:Introduction}

Despite the great advances that have been made, understanding how
galaxies form and evolve with time remains an ongoing challenge in modern
astronomy. Galaxies are complex, with the relevant physics
occurring on size scales ranging from individual stars and black holes
through to galaxy
clusters. A multi-wavelength approach to the problem is clearly
required, as the different processes and evolutionary stages are characterised by
emission mechanisms spanning the full electromagnetic spectrum.

Although observations of the stellar component of galaxies are
routinely made for large samples of objects over significant sky areas
(e.g. Sloan Digital Sky Survey, SDSS, \citealt{York2000}) to high redshift
(e.g. VIMOS VLT Deep Survey, VVDS, \citealt{LeFevre2005}),
observations of the gas in galaxies lags behind. As cold gas serves as the reservoir
from which stars form, the non-trivial relationship between the gas
and stellar content of galaxies is fundamental to understanding their
evolution. Observations of both the stellar and gaseous components of
galaxies are required to illuminate the interplay between the two.

Targeted observations of small numbers of individual galaxies provide
important insight. The \HI\ Nearby Galaxy Survey (THINGS, \citealt{Walter2008})
focuses on relatively local, highly resolved galaxies, enabling discrete
sites of star formation activity to be individually assessed based on
observations of the neutral and molecular gas 
(\citealt{Bigiel2008}). From these specific studies, the general
relations, observed on galaxy scales, can be better understood (such
as the Kennicutt-Schmidt law for star formation;
\citealt{Kennicutt1998}, or \citealt{Bigiel2012}). 

Large surveys, although insensitive to the details extracted from
targeted observations, reveal global trends governing the general
population. Often, statistical studies of such surveys provide
insight into processes too subtle to discern from
individual objects. At optical wavelengths, the SDSS has been
particularly effective at uncovering 
global trends in the stellar properties of galaxies, such as the red
sequence and blue cloud galaxy bimodality seen by \citet{Baldry2004},
for example.

The Square Kilometre Array (SKA, \citealt{Carilli2004}) will
revolutionize the study of extragalactic gas reservoirs, enabling
neutral hydrogen (\HI) to be observed to $z\sim 1$ over a 
significant fraction of the sky, bringing \HI\ observations more in
line with optical surveys, both in terms of sensitivity and area coverage. The SKA
pathfinder and precursor instruments APERTIF (\citealt{Verheijen2008}), MeerKAT
(\citealt{Jonas2009}) and the Australian 
Square Kilometre Array Pathfinder (ASKAP, \citealt{Johnston2008}),
currently under construction will already represent great advances
themselves, due to innovative receiver design and expanded frequency range.

Along with advancing instrumentation, models of galaxy evolution are
becoming increasingly sophisticated, with improved mass resolution and
more physical prescriptions. In modern semi-analytic models, galaxy
baryonic mass is partitioned into 
gas and stars, while the gas is further subdivided into atomic and
molecular components using a variety of methods
(\citealt{Obreschkow2009a}, \citealt{Lagos2011}, 
\citealt{Popping2012}). The interplay
between gas and stars is then governed by prescriptions based on local
conditions. Smoothed particle hydrodynamic simulations, such as those of
\citet{Dave2013}, have the resolution required to follow the detailed
dynamics of gas and stars on sub-galactic scales, but have
insufficient volume to model statistically representative samples of galaxies.
These simulations are invaluable, as they allow us to 
explore the interactions between the various galactic components,
including the influence of the dark matter haloes (\citealt{Popping2014}).

In the current work, we focus on the dependence of \HI\ mass (\MHI) on 
stellar mass within the \HI-selected galaxy population, and we exploit
this framework to compare observations with simulations. The relation between the
stellar and gaseous components in galaxies appears straightforward at
first glance, but the details provide insight into the history of both
gas accretion and star formation, and reveals the potential for
further star formation. Differences between the observed 
and simulated relations offer the opportunity to explore missing
physics or other factors driving galaxy evolution currently not
incorporated in the simulations. The data we use are described in 
Section~\ref{sec:data}, and presented in
Section~\ref{sec:MHIvsMstar}. A discussion of the 
results and suggestions for further lines of investigation are in
Section~\ref{sec:discussion}. Concordance cosmology with $H_{0} = 70$
km s$^{-1}$ Mpc$^{-1}$ (thus $h\equiv H_{0}$/[100 km s$^{-1}$
Mpc$^{-1}$]$=0.7$), $\Omega_{m} = 0.3$, $\Omega_{\Lambda} = 0.7$ is
assumed when computing masses and luminosities.

\section{\HI\ and Optical Data}\label{sec:data}

The \HI\ data for the current work come from the Arecibo Legacy Fast
ALFA survey (ALFALFA, \citealt{Giovanelli2005}), specifically the
$\alpha$.40 \HI\ source catalogue from \citet{Haynes2011}, which
covers $\sim$2800 deg$^2$, or 40~per~cent of the final survey
area. ALFALFA is a flux-limited survey, with sufficient sensitivity to
detect galaxies with \MHI $=3\times 10^7$\,\Msun\ at the distance of the
Virgo Cluster (corresponding to $\sim$0.47 Jy\ \kms), while the spectral
range allows observations of 
galaxies to $z=0.06$. \HI\ profile widths, recession 
velocities, distances and \HI\ masses are derived by the
ALFALFA team for each of the 15\,855 objects in the catalogue. Of
these, 15\,041 are confident \HI\ detections (\HI\ code 1 or 2 in the
ALFALFA catalogue), with the remainder being
primarily high velocity clouds, not at extragalactic distances.

The $\alpha$.40 catalogue has been carefully crossmatched to the SDSS Data Release 7 (DR7,
\citealt{Abazajian2009}) by members of the ALFALFA team to identify
the most likely optical counterparts for the \HI\ detections. The crossmatched
catalogues provide the SDSS \texttt{objID} and \texttt{specObjID}
identifiers of the optical galaxies (see \citealt{Haynes2011}
for a full description of the catalogue). 11740 of the 15855 ALFALFA
galaxies have confident \HI\ detections and unambiguous, clean optical
counterparts in the SDSS photometry (optical counterpart code I in the
ALFALFA catalogue). We refer to this subset as the
ALFALFA--SDSS sample. \HI-detected galaxies lying outside the SDSS
footprint (2312 objects) are the primary cause for ALFALFA galaxies
having no optical counterpart. 

A collaboration of researchers at the Max Planck Institute
for Astrophysics (MPA) and the Johns Hopkins University (JHU) have produced
a value-added galaxy catalogue (hereinafter referred to as the
\textit{MPA--JHU catalogue}), which provides additional metrics
derived from an independent analysis of the SDSS DR7 galaxy photometry
and spectra. Objects must have an SDSS spectrum to appear in the
MPA--JHU catalogue. Details of the analyses can be found
in \citet{Tremonti2004} and \citet{Brinchmann2004}, or at the website hosting the
catalogues\footnote{\url{http://www.strw.leidenuniv.nl/~jarle/SDSS}}. 
Of the ALFALFA--SDSS sample, 9471 have entries in the MPA--JHU
catalogue.

From the MPA--JHU catalogue, we extract measures of the stellar masses
(\Mstellar), derived by fitting the five optical SDSS photometry bands
with a grid of updated stellar population synthesis models based on
\citet{Bruzual2003}. The MPA--JHU group compared these masses with
those from \citet{Kauffmann2003} derived from spectral features and
found good agreement over all masses. At the low mass end, there is
also good agreement between the MPA--JHU stellar masses and those
computed from UV--optical photometry by \citet{Huang2012a} for the
galaxies contained in both samples. Of the
9471 galaxies in the MPA--JHU catalogue, 9153 have reliable stellar
mass estimates, which we refer to as the spectroscopic ALFALFA--SDSS
sample, and use in the following sections.

We have chosen to use ALFALFA for the current study, instead of, for
example, the \HI\ Parkes All-Sky Survey (HIPASS, \citealt{Barnes2001}, 
\citealt{Meyer2004}) due to the corresponding availability of homogeneous, high
quality ancillary photometry and spectroscopy from SDSS.


\section{\HI\ mass \textit{vs} stellar mass}\label{sec:MHIvsMstar}

\begin{figure}
\resizebox{\hsize}{!}{\includegraphics{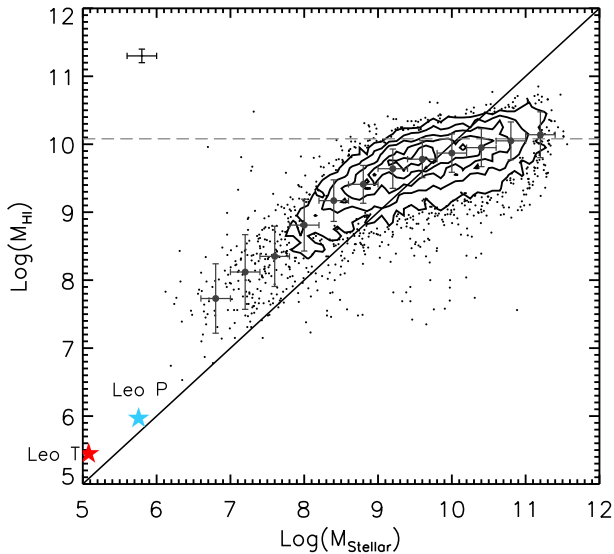}}
\caption{The relation between \MHI\ and \Mstellar, in units of \Msun, from the 9153
  \HI-selected ALFALFA galaxies with SDSS spectra and stellar masses
  from the MPA--JHU catalogue. The diagonal black solid line marks
  the 1-to-1 relation, while the horizontal grey dashed line indicates
  the approximate 50~per~cent completeness limit for profiles of width
  150 \kms\ at $z=0.06$, the highest redshift probed. The dark grey
  points with error bars
  show the median and 1-$\sigma$ values from the first three columns
  of Table~\ref{tab:medians}. The blue and red stars denote the very low mass 
  Leo P and Leo T dwarf galaxies. Typical uncertainties on \MHI\ and \Mstellar\ are 10 and
  20~per~cent, respectively, indicated by the example point in the top left
  corner with error bars. The few points at low stellar mass and \MHI $>10^9$\Msun\
  are pairs of galaxies within one ALFALFA
  beam at similar redshift, causing confusion with their \HI\
  profiles. The contour levels are at 5, 15, 30, 45, and 60 points
  computed on a grid with 0.1 intervals in log(\MHI) and log(\Mstellar).}
\label{fig:MHI_Mstellar}
\end{figure}

In Fig.~\ref{fig:MHI_Mstellar}, we show \MHI\ plotted as a function of 
\Mstellar\ for the spectroscopic ALFALFA--SDSS galaxy sample, which spans more
than five orders of magnitude in stellar mass, and nearly four orders of
magnitude in \HI\ mass. The diagonal black line marks
the 1-to-1 relation of equal \HI\ and stellar masses. The points show
a clear correlation, with \MHI\ increasing as a function of \Mstellar, and a
break in the slope of the relation at \Mstellar$\sim 10^9$\,\Msun. Similar plots,
such as Figure 2a of \citet{Huang2012b}, also show this behaviour. The
median and 1-$\sigma$ values for the data in
Fig.~\ref{fig:MHI_Mstellar} in bins of stellar mass are tabulated in Table~\ref{tab:medians}. 

It is known that more massive galaxies have higher molecular gas
to neutral gas ratios (\Htwo/\HI, \citealt{Blitz2006},
\citealt{Leroy2008}). In order to maintain a continuous 1-to-1
relation from low to high masses, the mass of \Htwo\ required at the
high mass end is of the order of the mass of \HI, and galaxies with
such large amounts of molecular gas are not observed in the disk
galaxies which dominate the ALFALFA sample.

At the high stellar mass end, the galaxies' \HI\ content is weakly
dependent on stellar mass. These galaxies are
dominated by metal-enriched disks, with gas-phase oxygen metallicities,
12+log(O/H), compiled by the MPA--JHU group, reaching solar values and
above. This is consistent with the conclusion reached in
\citet{Huang2012b}, which shows that the ALFALFA population primarily
exists within the blue cloud. The number counts of \HI-detected early-type
and spiral galaxies within the ATLAS$^{3D}$ project also show that
disks dominate at large masses (\citealt{Serra2012}). At lower stellar masses, 
below the break at \Mstellar$\sim 10^9$\,\Msun, the galaxies have
relatively high \HI\ fractions, are metal poor, and as discussed in
Section~\ref{sec:lowmass}, tend to have irregular morphology.

The blue star in the bottom left corner of Fig.~\ref{fig:MHI_Mstellar}
is the Leo P dwarf galaxy, while the red star at even lower mass is
the Leo T dwarf galaxy, indicating that the linear relation between
\MHI\ and \Mstellar\ at \MHI\ $<10^9$\Msun\ appears to continue to
very low neutral gas and 
stellar masses. The strength of the ALFALFA survey is clearly illustrated
here: its high sensitivity allows us to explore the trend over a wide range in
\HI\ masses. Observing only the most \HI-massive galaxies
would lead to the incorrect assumption that \HI\ mass is only very weakly
dependent on stellar mass.

\begin{table}
\centering
  \caption{\label{tab:medians} Median and 1-$\sigma$ values for the
    data shown in Fig.~\ref{fig:MHI_Mstellar} and
    Fig.~\ref{fig:nospecMLR}. The first column lists the centre of the
    log(\Mstellar)=0.4 width bins. The second and third columns are
    derived only from galaxies with SDSS spectra, whereas the fourth
    and fifth columns include galaxies without a spectrum. Note that the 
    width of the distribution of points is not symmetric about the median, thus the
    1-$\sigma$ values are provided only for guidance.}
\begin{tabular}{ccccc} \hline
log(\Mstellar) & Median log(\MHI) & 1-$\sigma$ & Median log(\MHI) & 1-$\sigma$\\ 
\Msun & \Msun & \Msun & \Msun & \Msun \\ \hline
6.8 & 7.73 & 0.51 & 8.13 & 0.67 \\
7.2 & 8.12 & 0.55 & 8.44 & 0.63 \\
7.6 & 8.35 & 0.44 & 8.68 & 0.59 \\
8.0 & 8.81 & 0.38 & 8.99 & 0.45 \\
8.4 & 9.17 & 0.31 & 9.23 & 0.34 \\
8.8 & 9.41 & 0.28 & 9.43 & 0.30 \\
9.2 & 9.64 & 0.29 & 9.64 & 0.30 \\
9.6 & 9.78 & 0.27 & 9.77 & 0.28 \\
10.0 & 9.87 & 0.28 & 9.86 & 0.29 \\
10.4 & 9.95 & 0.28 & 9.95 & 0.29 \\
10.8 & 10.05 & 0.28 & 10.04 & 0.29 \\
11.2 & 10.14 & 0.34 & 10.14 & 0.33 \\ \hline
\end{tabular}
\end{table}

\subsection{\HI\ and optical completeness}\label{subsec:complete}

It is important to recall that our galaxy sample is \HI\
flux-limited. The completeness of ALFALFA as a function of \HI\ flux
and profile width is discussed in \citet{Haynes2011}. Assuming no
self-absorption, the total \HI\ flux received from a galaxy depends
only on the \HI\ mass, and the luminosity distance to the galaxy, $D_L$, in Mpc: 

\begin{equation}
M_{\mathrm{HI}} = 2.356\times10^5\, D_{L}^2\,(1+z)^{-1} \int S_{v} \,d v \\
\label{eq:HImass}
\end{equation}

\noindent where \MHI\ is in solar masses, and the integral is the
total flux in Jy \kms. The $(1+z)$ factor accounts for the difference
between the observed and rest-frame profile width. Due to the distance
dependence, galaxies with low \MHI\ are only detected at the lowest
redshifts, and not the full survey volume. This completeness leaves the
area towards the bottom right of the locus of points in the figure
underpopulated, as these galaxies are both relatively gas-poor, and their
volume density is low. The horizontal grey dashed line in
Fig.~\ref{fig:MHI_Mstellar} indicates the 50~per~cent completeness of
ALFALFA at $z=0.06$ for a galaxy with width $W_{50}=150$\,\kms, thus
galaxies with masses above this are visible over the full survey volume.

In addition to the \HI\ selection, galaxies must also have an SDSS 
spectrum to appear in Fig.~\ref{fig:MHI_Mstellar}. A galaxy can be
without a spectrum either because it is fainter than the spectroscopic
limit of $r=17.77$, or was simply not assigned a fibre by the tiling
algorithm. Thus, galaxies without spectra span the full range of
stellar masses, but galaxies with spectra almost all have $r\le
17.77$. At $z=0.06$, the ALFALFA--SDSS galaxies with $r\sim 17.7$ have
stellar masses of log(\Mstellar) $\sim 9.3$. Above this stellar mass, galaxies are brighter
than $r=17.77$ over the full survey volume. 

For the galaxies detected in ALFALFA that do not have spectra, and thus
stellar masses from the MPA--JHU catalogue, we can estimate their
stellar masses using scaling relations from the observed optical 
magnitudes. A correlation is seen between log(\Mstellar/\Lstellar$_i$)
and ($g-r$) in \citet{Bell2003}, where \Lstellar$_i$\ is the stellar luminosity
derived from the SDSS $i$-band magnitude. We assume that the
\HI-selected galaxies without optical spectra are a 
subset of the full galaxy sample and follow the same scaling relations.

\begin{figure}
\resizebox{\hsize}{!}{\includegraphics{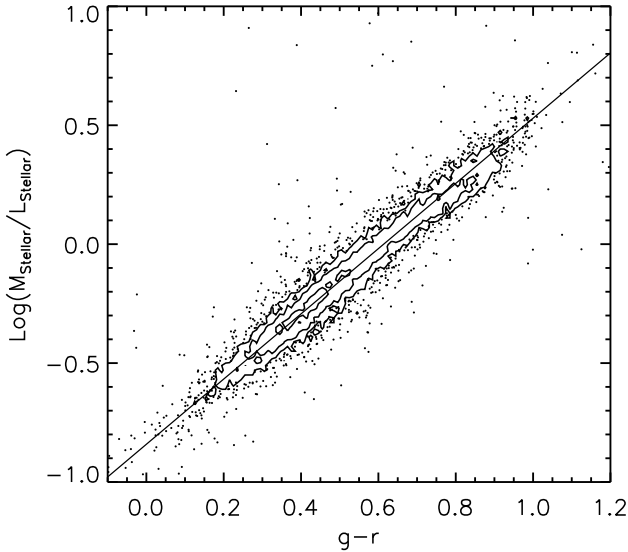}}
\caption{The log(\Mstellar/\Lstellar$_i$) \textit{vs} ($g-r$) relation derived
  from the spectroscopic ALFALFA--SDSS galaxies, using \Mstellar\ from
  the MPA--JHU catalogue, and \Lstellar$_i$ from the galaxies' SDSS
  $i-$band apparent magnitudes. The solid black line shows the fit to
  the data points.}
\label{fig:MLR}
\end{figure}

\begin{figure}
\resizebox{\hsize}{!}{\includegraphics{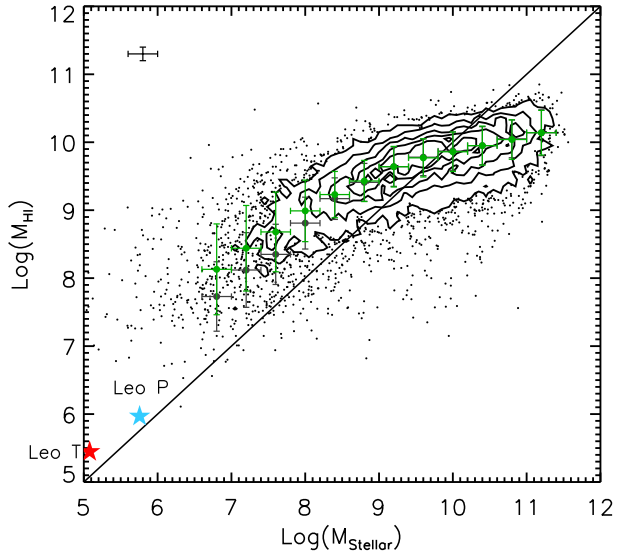}}
\caption{Similar to Fig.~\ref{fig:MHI_Mstellar}, but including ALFALFA
galaxies that have SDSS imaging but do not have an SDSS spectrum.
The stellar masses for these galaxies are computed from
the relation derived in Fig.~\ref{fig:MLR}. The additional galaxies
lessen the severity of, but do not eliminate, the break in the
envelope, and the dispersion increases at low masses. The median
values for the full sample, from the fourth and fifth columns of
Table~\ref{tab:medians}, are shown as green points with error bars, while the median
values for the spectroscopic sample are in grey as in Fig.~\ref{fig:MHI_Mstellar}. The additional
galaxies span the full range of stellar masses.}
\label{fig:nospecMLR}
\end{figure}

Using the galaxies in the spectroscopic
ALFALFA--SDSS sample with known stellar masses, we derive new
coefficients for the log(\Mstellar/\Lstellar$_i$) -- ($g-r$) correlation, which reflect the
biased nature of our \HI-selected galaxy sample. We find
log(\Mstellar/\Lstellar$_i$) = -0.84 + 1.37($g-r$) for the points shown in
Fig.~\ref{fig:MLR}, and can then estimate the
stellar masses for galaxies without spectra using the updated coefficients.
The additional points are included in Fig.~\ref{fig:nospecMLR}, and
lie in the same regions on the \MHI--\Mstellar\ plane as the 
galaxies with MPA--JHU stellar masses, thus the shape of the upper
envelope is not affected by the optical flux limit.

To populate the top left quadrant of Fig.~\ref{fig:MHI_Mstellar} and
Fig.~\ref{fig:nospecMLR}, at low stellar 
mass and high \HI\ mass, galaxies would require \MHI/\Mstellar$\sim
100$. If these galaxies existed in large numbers, they would be
sufficiently \HI-rich such that ALFALFA would detect them. Their low
stellar masses, however, put them near the magnitude limit of
the SDSS photometry, $r\sim 22$, which at $z=0.06$ corresponds to
\Mstellar\ $<10^7$\Msun. There are a small number (less than one per~cent
of the $\alpha$.40 catalogue) of \HI-bright, optically very faint, galaxies
within ALFALFA, but they are indeed very rare
(\citealt{Haynes2011}). Thus, the galaxies shown here trace a real
upper envelope of \HI\ mass fraction, \fHI\ $\equiv$ \MHI/\Mstellar.

\subsection{Models of \HI\ mass \textit{vs} stellar mass}\label{sec:scubed}

A number of models exist which incorporate a large range of
physical processes for the formation and evolution of
galaxies, including accretion, expulsion and ionisation of gas, and conversion of
gas into stars. Relevant for this work is the partitioning of gas into
neutral, ionized, and molecular components. Modern semi-analytic
models are able to match the $z=0$ stellar and/or \HI\ 
mass functions, either by construction or as a consistency
check. However, none have performed a detailed comparison to the
two-dimensional \MHI--\Mstellar\ distribution over the full range of
galaxy masses shown in Fig.~\ref{fig:MHI_Mstellar}. This comparison tests
whether the \HI\ gas is located in galaxies of the correct stellar
mass. 

We have chosen to make a first comparison with galaxies extracted from
the SKA Simulated Skies (S$^3$)
simulations\footnote{http://s-cubed.physics.ox.ac.uk/}
(\citealt{Obreschkow2009a}, \citealt{Obreschkow2009b}). The
simulations build on the dark matter framework of the Millenium
Simulation (\citealt{Springel2005}) and the semi-analytic galaxy formation
prescriptions of \citet{DeLucia2007}. Details
regarding how \MHI\ is computed for the simulated galaxies can be
found in \citet{Obreschkow2009a}. The
S$^3$ galaxies are in reasonably good agreement with the HIPASS \HI\ mass
function at \MHI\ $>10^8$\,\Msun\ and the Tully-Fisher relations at $z=0$. 

\begin{figure}
\resizebox{\hsize}{!}{\includegraphics{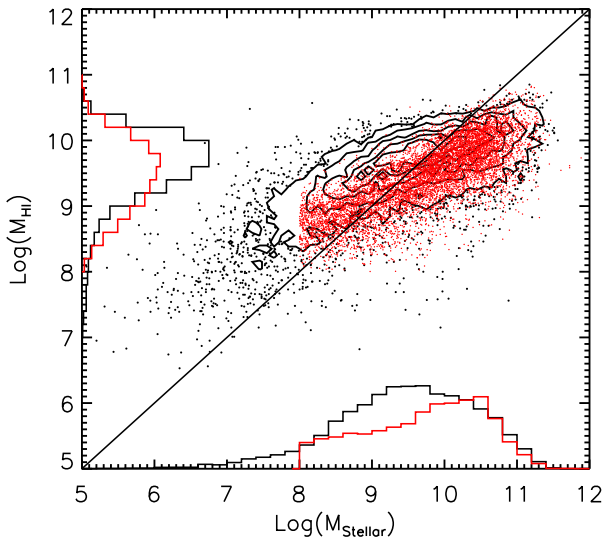}}
\caption{\MHI\ vs \Mstellar\ for the ALFALFA--SDSS galaxies, including
  galaxies without spectra (black
  points and contours), and galaxies extracted from a volume within the S$^3$
  simulations approximating the ALFALFA $\alpha$.40 volume (red
  points). The S$^3$ simulated galaxies lack a break in the relation at \Mstellar$\sim
  10^9$\,\Msun. The histograms along the $x$-axis show the
  stellar mass distributions, and along the $y$-axis show the \HI\ mass
  distributions, for the data and simulations in black and red, respectively.}
\label{fig:paper_sims}
\end{figure}

Fig.~\ref{fig:paper_sims} compares the ALFALFA--SDSS
galaxies (black points and contours) to
the S$^3$ simulated galaxies (red points). The S$^3$ galaxies are
extracted from a simulation volume approximating the ALFALFA
$\alpha$.40 volume, also incorporating the 50~per~cent completeness limits from
Equations 4 and 5 in \citet{Haynes2011}. The simulated galaxies are
only shown for \Mstellar\ and \MHI\ $>10^{8}$\,\Msun, as the underlying dark matter
simulation does not resolve smaller galaxies. The simulated galaxies
overlap with the data at the highest \HI\ and stellar mass, but the
locus of points does not follow the ALFALFA--SDSS data traced by the
black contours. The difference is largest near the break in
the observed galaxy distribution. While the simulations are restricted to \MHI
$>10^{8}$\,\Msun, there is some evidence that haloes hosting galaxies
with \Mstellar $<4\times 10^{9}$\,\Msun\ have only existed for a few
simulation timesteps, and their properties are not yet fully converged
(\citealt{Obreschkow2009a}). However, differences between the
simulations and data already appear at \Mstellar$=10^{10}$\,\Msun. 

The black and red
histograms along the $x$- and $y$-axis in Fig.~\ref{fig:paper_sims} show
the distributions of stellar and \HI\ masses for the data and
simulation, respectively. The simulation has too many high stellar
mass (\Mstellar$>10^{10}$\,\Msun) galaxies in the \HI-selected sample,
and too few galaxies with 
$10^9<$\Mstellar$<10^{10}$\,\Msun. The simulated galaxies with
moderate stellar masses ($10^8<$\Mstellar$<M^{10}$\,\Msun) are also
too \HI-poor with respect to the observed data. This suggests that
model galaxies of present-day stellar masses around \Mstellar$\sim
10^9$\,\Msun\ recently had too little cold gas accretion or star formation
efficiency was too high. This, in turn, could indicate missing physics in the
model, such as a missing prescription for ram pressure stripping or
neglected non-linearities in the partitioning between ionised, atomic
and molecular gas.

\begin{figure}
\resizebox{\hsize}{!}{\includegraphics{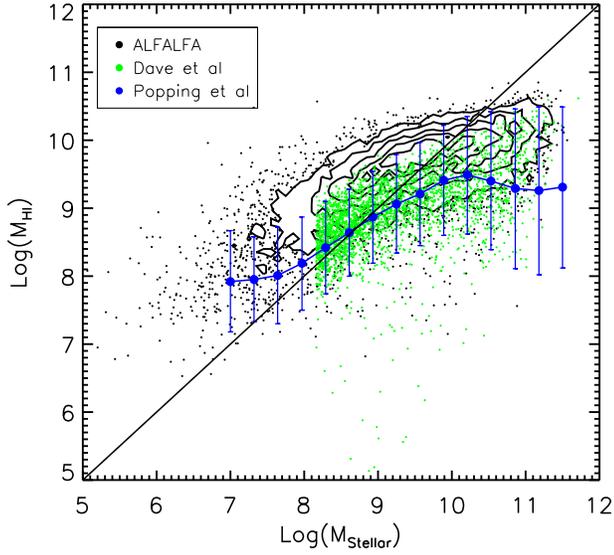}}
\caption{\MHI\ vs \Mstellar\ for the ALFALFA--SDSS galaxies, including
  galaxies without spectra (black 
  points and contours), and simulated galaxies from \citet{Dave2013}
  (green dots) and \citet{Popping2014} (blue dots with 2-$\sigma$
  error bars). The Dav{\'e} 2013 galaxies follow a similar trend to the
  S$^3$ galaxies, while the Popping 2014 simulations do show a break
  in the relation, albeit at a higher stellar mass than the ALFALFA galaxies.}
\label{fig:mikagergo}
\end{figure}

A further comparison can be made with the smoothed particle \textit{N}-body + 
hydrodynamics simulations from \citet{Dave2013}. Galaxies
extracted from their smaller simulation volume of $\sim$45 Mpc on a
side with the `ezw' outflow model show qualitatively similar behaviour
to the S$^3$ galaxies, seen as green dots in Fig.~\ref{fig:mikagergo}. Also shown in
the figure is the mean trend of \MHI\ as a function of \Mstellar\ for
the galaxy population simulated by \citet{Popping2014} (blue points
with 2-$\sigma$ error bars). These simulations are tuned to match the
$z=0$ stellar mass function, and the only selection criterion imposed
here was that the galaxies are disk-dominated, with
$M_{bulge}/M_{total}<0.4$ to reflect the bias of the ALFALFA galaxies
toward disks. As no restrictions on the observed \HI\
flux is imposed on the galaxy sample extracted, relatively \HI-poor
galaxies may be biasing the mean \MHI\ toward lower values. 
Unlike the S$^3$ model, the simulations and models by
\citet{Dave2013} and \citet{Popping2014} have not yet been turned into
mock skies. Therefore, it is currently impossible to extract virtual
samples that match the ALFALFA+SDSS selection from these models.
Although a hint of a break in the relation is seen from the Popping et
al. simulations, it is at \Mstellar$\sim 10^{10}$\Msun, much higher
than seen in the data. 

The different behaviour of the galaxies at low and high stellar masses
with respect to their \HI\ content is a challenge for simulations. As
seen in Fig.~\ref{fig:paper_sims} and Fig.~\ref{fig:mikagergo}, none
of the simulated galaxies reproduce
the two regimes, implying a mass-dependent ingredient is
missing in the prescriptions regulating the \HI\ fraction in galaxies.
In fact, galaxies are not scale-free with respect to their \HI\
fraction. Whether the transition is governed by stellar mass, \HI\
mass or dark matter halo properties is unclear. 


\section{Discussion}\label{sec:discussion}

Considering the complexity of the interactions between gas and
stars, and the diversity of galactic environments, it is remarkable that a
distinct upper envelope of \HI\ mass at a given stellar mass exists at
all. One might naively expect that stochastic processes such as
environment-dependent gas accretion, different types of outflows, 
galaxy mergers and subsequent triggered star formation prohibit such a
relation. In the following discussion 
we attempt to explain the upper \MHI\ envelope, and the break in the
slope seen in the data at \Mstellar$\sim 10^9$\,\Msun. 

The buildup of \HI\ mass and stellar mass is different from
other mass relations, such as the correlation between the mass of
central black holes and their host spheroids. In the former, stellar
mass is built at the expense of the \HI\ reservoir. A major merger of
two gas-rich disks results in a triggered burst of star formation,
resulting in a massive stellar system with only a fraction of the
resulting total mass retained as \HI. For the latter mass relation,
both the black hole and spheroid can continue to grow, provided there
is a source of fuel. 

It is important to remember that the objects appearing in
the current work are \HI-selected, and represent only
galaxies with a significant mass of neutral hydrogen.
The figures show a snapshot during the phase when they
are \HI-rich, and does not represent an evolutionary locus for galaxies to
follow. Galaxies below the mass sensitivity of the
ALFALFA survey galaxy lie below the locus of
points.

\subsection{What causes the upper limit of \fHI?}\label{sec:envelope}

From Fig.~\ref{fig:MHI_Mstellar}, it is clear that there is a stellar mass-dependent upper
limit on galaxies' \HI-to-stellar mass ratio. In the following, we
investigate possible causes of this upper limit. 

\subsubsection{Dark matter halo spin parameter}\label{sec:highmass}

While discussing the properties of the ALFALFA galaxies,
\citet{Huang2012b} make a convincing argument for galaxies with higher 
\HI\ fractions at a given stellar mass existing in dark matter haloes with larger spin
parameters, and thus have more extended disks (see their Figure 14b). This
is consistent with work from \citet{Boissier2000}, who find that
haloes with large spin parameters host more extended \HI\ disks, along
with a more quiescent star formation history. A similar 
link between halo spin parameter and \HI\ disk size is
found in \citet{Obreschkow2009a}. Also note that there is a tight positive
correlation between the mass of an \HI\ disk and the \HI\ disk radius, as
seen in Figure 7 of \citet{Verheijen2001}, showing that as disks
become more massive, they grow larger. This is a natural
consequence of the upper limit on \HI\ density found by
\citet{Bigiel2008}, for example, beyond which \HI\ rapidly becomes molecular.

Following the method outlined in \citet{Huang2012b}, we compute
the dark matter halo spin parameters for the ALFALFA--SDSS
galaxies. From \citet{Boissier2000}, we define the dimensionless halo spin
parameter, $\lambda$, as:

\begin{equation}
\lambda = \frac{\sqrt{2} V_{rot}^2R_d}{GM_{halo}}\\
\label{eq:lambda}
\end{equation}

\noindent where $V_{rot}$ is the rotation velocity of the disk, $R_d$ is the
scale length of the stellar disk, assumed to have an exponentially
declining surface density, and $M_{halo}$ is the halo mass. 
Equation \ref{eq:lambda} is only appropriate for mass
distributed in regular disks, which the vast majority of galaxies at
the high stellar mass end are. Further, Equation~\ref{eq:lambda}
assumes that an equal fraction of mass and specific angular momentum
is transferred from the halo to the disk, i.e. $m_{disk}/j_{disk}=1$ \citep{Boissier2000}.
We determine the galaxies' inclination, $i$, using the $r$-band axis ratio 
$b/a$ from the SDSS imaging, and assume an intrinsic axis ratio of
$q_0=0.2$ for reasons outlined in \citet{Huang2012b}:

\begin{equation*}
\mathrm{cos}^2 i = \frac{(b/a)^2 - q_0^2}{1-q_0^2}\\
\end{equation*}

\noindent From this we can convert the \HI\ profile width from the 
ALFALFA catalogue, $W_{50}$, into the rotation velocity,
$V_{rot}$, via $V_{rot}=(W_{50}/2)/\mathrm{sin}(i)$. $R_d$ is extracted from
the SDSS database as the $r$-band exponential fit scale radius, converted to kpc. 
We convert the $V_{rot}$ to the rotation velocity of the associated
dark matter halo, $V_{halo}$, via the relation found in
\citet{Papastergis2011}. Finally, the virial mass of the dark matter
halo is determined from the relation derived from simulations from
\citet{Klypin2011}, where \mbox{$V_{halo}=2.8\times
10^{-2}(M_{halo}h)^{0.316}$}. 

\begin{figure}
\resizebox{\hsize}{!}{\includegraphics{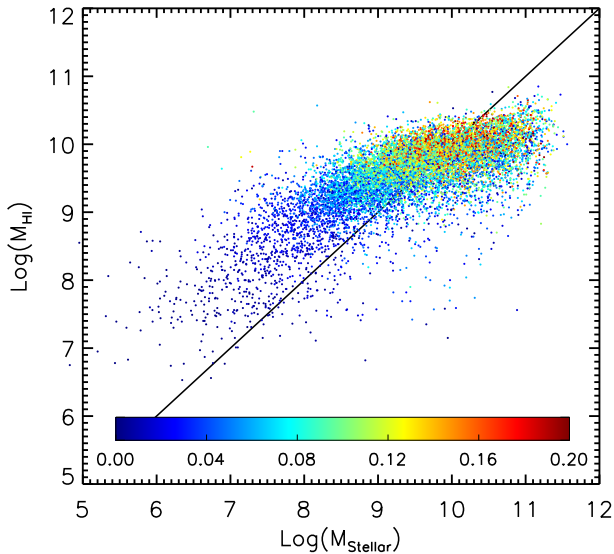}}
\caption{\MHI\ vs \Mstellar\ for the ALFALFA--SDSS
  galaxies, including galaxies without spectra, colour coded by their
  dark matter halo spin parameter. At high masses, 
  galaxies with the larger \fHI\ are also located within haloes with
  the largest spin parameter. At low masses, all haloes have low
  spin parameters, even though these galaxies have the highest \HI\ fractions.}
\label{fig:MHI_spin}
\end{figure}

Empirically, we find a positive correlation between \HI\ fraction and
halo spin: at any fixed stellar mass \Mstellar$> 10^8$\,\Msun,
galaxies with higher \HI\ fractions sit in haloes with higher spin parameters. This
finding is consistent with the picture that a high spin can stabilise
a large \HI\ disk to prevent it from clumping and forming stars. A
natural consequence of this finding is that very high \HI\ mass
fractions, or equivalently, very high \HI\ masses at a given stellar
mass, would require very high spin parameters. However, 
theoretically, the spin is limited by the amount of infall and tidal
torque haloes can experience during the proto-galactic growth. Explicit
numerical N-body simulations of CDM-haloes by \citet{Knebe2008} find
an upper limit around $\lambda \sim 0.2$, independent of halo mass. This
value is consistent with the highest spin parameters measured in this
study. Combining the theoretical prediction of an upper limit in the
spin parameter with the empirical correlation between spin and \HI\
fraction provides a plausible explanation of the \HI\ fraction in
galaxies: The maximum \HI\ fraction is set by the upper limit in the
spin parameter. 

As \HI\ observations only provide a measure of the neutral gas
content, the upper limit on \HI\ mass could also be due to ionization of the
neutral gas. \citet{Portas2012} find that the \HI\ density in the
THINGS galaxies they studied is too high for the background ionizing
field to be responsible for the sharp truncation of \HI\ disks,
indicating that the maximum \HI\ mass seen here is not due to
increasingly large quantities of ionized gas, but is rather an accurate 
reflection of an upper limit of the gas content of galaxies. Galaxy
interactions, or ram-pressure stripping, are also able to 
truncate \HI\ disks. However, as \HI-rich galaxies are generally found
in low-to-moderate density environments (\citealt{Haynes2011},
\citealt{Huang2012b}), neither of these effects dominate the \HI\
properties of the galaxies discussed here.

Investigating the dark matter halo spin parameter upper limit, and the
formation of large galaxies in haloes with high spin parameter, will
help explain the behaviour of the most \HI-massive galaxies and the
corresponding upper limit on \HI\ mass. At these high stellar masses,
the galaxies are regular, massive disks, which are resistant to small
perturbations, and star formation can proceed efficiently. The
\HI ghMass project (\citealt{Huang2014}) aims to study in detail  
the most \HI-rich, massive galaxies. Preliminary results for two galaxies from their
sample \citep{Hallenbeck2014} indicate that one, UGC 12506, indeed
has a large, extended \HI\ disk and is located in a rapidly spinning dark matter halo, consistent
with the general trend found here. The other, UGC 9037, has a more
centrally concentrated \HI\ distribution, with non-circular flows
within the disk. Fairly regular optical and \HI\ morphology argues against a
recent major merger, rather suggesting accretion from a gaseous
halo. Similar analyses of their full sample will illuminate the
relative importance of dark matter halo spin properties as compared to 
environmental or morphological factors for the high mass galaxy population.

\subsubsection{Galaxy morphology}\label{sec:lowmass}

At the low stellar mass end (\Mstellar $< 10^9$\Msun), the upper
envelope is not delimited by high halo spin parameter, even though
these galaxies have the largest \HI\ fractions of all the
galaxies. Therefore, some other mechanism must 
restrict the \HI\ fraction. To test whether environmental effects set
an upper limit in \fHI\ in these low-mass galaxies, we employ the morphological
information derived from the Galaxy Zoo project
(\citealt{Lintott2008}), which can separate the ALFALFA--SDSS galaxies
into `spirals', `mergers', or `unknown' morphologies, based on human
votes. In practice, it is difficult to distinguish between `mergers' and
`unknowns', thus we simply require that the galaxies have low
probability of being `spirals', $p_{spiral}<0.4$. We refer
to the resulting galaxies as irregular. The following is not
sensitive to the exact restrictions on the galaxy type percentages.

At stellar masses below $10^9$\,\Msun, the upper envelope of \HI\
fraction is heavily populated by irregular galaxies. As the galaxies
are low mass, their morphology is easily disrupted. The number of
irregular galaxies then drops significantly between
$10^9$\Msun$<$\Mstellar$<10^{10}$\,\Msun, coinciding with the break in the 
relation at \Mstellar$\sim 10^9$\,\Msun. The predominance
of irregular galaxies at low masses may explain why 
there is no ordered gradient of halo spin parameters with \HI\
fraction.

The high \HI\ fractions at low stellar mass imply either inefficient star formation in
the past, or recent accretion of significant amounts of neutral
gas, or a combination of both. Using further information from the
MPA--JHU catalogue, we find that the galaxies with the highest \HI\
fractions also have the highest current specific star formation rates
of all the ALFALFA galaxies. The metallicity is also lowest for the highest \fHI\
galaxies, dropping below solar values to 12+log(O/H)$<8$. These two
ingredients, coupled with the 
galaxies' irregular morphology, favour recent gas accretion. These
galaxies could also have recently experienced an interaction 
with another gas-rich, low mass galaxy. Further \HI\ observations at
higher resolution, and deeper optical imaging, would be required to
distinguish between the interaction and accretion scenarios.

The link between \HI\ and molecular gas,
primarily \Htwo, is clearly an important stage in star formation, and
in disk galaxies is found to be related to pressure
(\citealt{Blitz2006}). For the low-mass galaxies, the lack of massive disks
may result in less efficient star formation. Testing this hypothesis 
requires simulations with sufficient resolution to follow gas
accretion onto galaxies with \Mstellar$<10^{7-8}$\,\Msun, coupled with
detailed observations of molecular gas of a sample of galaxies.

The transition between the spin-regulated and morphology-regulated
regimes, at \Mstellar$\sim 10^9$\,\Msun, also
coincides with the transition mass identified within the smoothed
particle hydrodynamical simulations of \citet{Keres2009}. They find,
at halo masses below $2-3 \times 10^{11}$\,\Msun, corresponding to
baryonic masses of $2-3 \times 10^{10}$\,\Msun, cold accretion is the dominant
accretion mode, whereas at masses greater than this, gas accretion 
switches to hot mode. This is consistent with the low stellar mass
galaxies having recently accreted a significant mass of cold gas.


\section{Summary}\label{sec:summary}

Despite the complicated interactions between galaxies and their
environments, along with energetic internal processes, there is a
distinct upper limit on the mass fraction of neutral hydrogen a galaxy
is able to support. This upper limit argues against the existence of a
significant population of the so-called `dark galaxies', which are
\HI-rich but host very few stars.

The maximum \HI\ fraction is different for low and high stellar mass galaxies, 
with a break occuring at \Mstellar$\sim 10^9$\,\Msun. At high stellar
masses, the dark matter halo spin 
parameter, as well as the \HI\ mass, both reach a maximum. At low
stellar masses, the most \HI-rich galaxies are morphologically irregular, 
have low metallicity and current high star formation efficiency,
indicating recent gas accretion. The transition between the two mass 
regimes corresponds to a change in the predominant morphology of the
galaxy populations, as well as the transition mass between hot and cold mode
accretion in simulations by \citet{Keres2009} at \Mbary $=2-3 \times
10^{10}$\,\Msun. Resolved \HI\ observations
of the low mass objects will provide further information regarding the
\HI\ content and how it is distributed in the galaxies. 

ALFALFA, covering several thousands of square degrees with the
spectral range and sensitivity to detect galaxies to $z=0.06$
represents a significant improvement over previous \HI\ studies. 
The SKA, pathfinder and precursor instruments APERTIF, MeerKAT and ASKAP mark the next
advance in \HI\ observations. The most significant gain will be due to
the greatly expanded spectral range, 
which will enable \HI\ in galaxies to be observed to cosmological
redshifts. Large numbers of galaxies with \MHI$<10^9$\,\Msun\ will be
observable at $z\sim 0.4$, populating the low mass end of
Fig.~\ref{fig:MHI_Mstellar}. The forthcoming \HI\ observations will be
supplemented by very deep radio continuum observations, from which
dust-free estimates of the star formation can be derived without the
need for supplementary optical imaging or spectroscopy. Observing
galaxies as a function of their environment and spanning significant
lookback time may allow us to view the buildup of the stellar mass
from the \HI\ reservoirs. 

Combining the next generation of \HI\ surveys with ever improving
simulations incorporating more sophisticated gas physics is sure to
further our understanding of the buildup of 
mass within galaxies and how they interact with their environment. 
The current discrepancies between the observations and simulations, in
particular the lack of a break in the envelope of \HI\ mass fraction, 
indicate physics not yet properly incorporated in the simulations.

\section*{Acknowledgments}

The Arecibo Observatory is operated by SRI International under a
cooperative agreement with the National Science Foundation
(AST-1100968), and in alliance with Ana G. Méndez-Universidad
Metropolitana, and the Universities Space Research Association. 
We acknowledge the work of the entire ALFALFA collaboration
team in observing, flagging, and extracting the catalogue of galaxies used in this
work. 

Funding for the SDSS and SDSS-II was provided by the Alfred P. Sloan
Foundation, the Participating Institutions, the National Science
Foundation, the U.S. Department of Energy, the National Aeronautics
and Space Administration, the Japanese Monbukagakusho, the Max Planck
Society, and the Higher Education Funding Council for England. The
SDSS was managed by the Astrophysical Research Consortium for the
Participating Institutions. The SDSS Web Site is
http://www.sdss.org/.

The SDSS is managed by the Astrophysical Research Consortium for the
Participating Institutions. The Participating Institutions are the
American Museum of Natural History, Astrophysical Institute Potsdam,
University of Basel, University of Cambridge, Case Western Reserve
University, University of Chicago, Drexel University, Fermilab, the
Institute for Advanced Study, the Japan Participation Group, Johns
Hopkins University, the Joint Institute for Nuclear Astrophysics, the
Kavli Institute for Particle Astrophysics and Cosmology, the Korean
Scientist Group, the Chinese Academy of Sciences (LAMOST), Los Alamos
National Laboratory, the Max-Planck-Institute for Astronomy (MPIA),
the Max-Planck-Institute for Astrophysics (MPA), New Mexico State
University, Ohio State University, University of Pittsburgh,
University of Portsmouth, Princeton University, the United States
Naval Observatory, and the University of Washington. 

NM wishes to acknowledge the South African SKA Project for
funding the postdoctoral fellowship position at the University of Cape
Town. KMH's research has been supported by the South African Research
Chairs Initiative (SARChI) of the Department of Science and Technology
(DST), the Square Kilometre Array South Africa (SKA SA), and the
National Research Foundation (NRF). 
We are grateful to Gerg\"o Popping and Mika Rafieferantsoa for
providing \HI\ and stellar masses from their simulations.
This work has benefitted from many useful discussions with a number of
people, including, but not limited to, Jacqueline van Gorkom, Martha Haynes, Gyula
Jozsa, Andrew Pontzen, Eric Wilcots, and Ian Heywood. We thank the
anonymous referee for helpful comments and quick response which improved this paper.
This research has made use of NASA's Astrophysics Data System.



\end{document}